\documentclass[twoside]{article}
\usepackage{fleqn,espcrc2}

\usepackage{graphicx}

\newcommand{\AmS}{{\protect\the\textfont2
  A\kern-.1667em\lower.5ex\hbox{M}\kern-.125emS}}

\title{Probing spin-charge separation 
using spin transport}

\author{Qimiao Si\\
Department of Physics, Rice University,
Houston, TX 77251-1892}
       
\begin{document}

\begin{abstract}
Pedagogical discussions are given on what constitutes a signature
of spin-charge separation. A proposal is outlined to probe spin-charge
separation in the normal state of the high $T_c$ cuprates using spin
transport. Specifically, the proposal is to compare the temperature
dependences of the spin resistivity and electrical resistivity:
Spin-charge separation will be manifested in the different temperature
dependences of these two resistivities. We also estimate the spin
diffusion length and spin relaxation time scales,
and we argue that it should be experimentally feasible to measure
the spin transport properties in the cuprates using the spin-injection
technique. The on-going spin-injection experiments in the cuprates
and related theoretical issues are also discussed.
\vspace{1pc}
\end{abstract}
\maketitle

\section{Introduction}

The normal state properties in the high temperature 
superconductors are anomalous in the context of 
the Fermi liquid theory. Two examples come from charge 
dynamics\cite{review}: The electrical resistivity
and inverse Hall angle have a linear and quadratic
temperature dependences, respectively. For the optimally
doped cuprates, such temperature dependences occur
all the way to temperatures as high as about 1000 K.
These anomalous properties suggest an unusual excitation
spectrum. Precisely what are the elementary excitations,
however, remains an open question.

One particular debate is whether or not spin-charge
separation exists. Since the initial suggestion\cite{Anderson},
considerable efforts have been devoted to interpret the
existing experimental data using spin-charge separation
based pictures. On the other hand, these
data have also been analyzed without invoking spin-charge
separation. Readers are referred to this proceedings for a 
snapshot of this continuing debate.

Theoretically, it is still not certain how spin-charge
separation arises in specific models in two dimensions.
The theoretical challenge is how to study in a controlled
fashion the non-perturbative effects of electron-electron 
interactions, as it is known that the Fermi liquid theory is
stable when interactions are treated perturbatively.
The situation is different from one dimension, where the phenomenon
of spin-charge separation was first discovered theoretically\cite{Voit},
as well as the opposite limit of large dimensions where metallic
states with spin-charge separation have also been shown to
occur in some specific models.

Instead of discussing these microscopic issues, here we address 
a phenomenological question: What would be (unambiguous)
experimental signatures of spin-charge separation? 
In the remainder of this paper, we elaborate on what constitutes
a signature of spin-charge separation, review a specific proposal
of using spin transport as such a probe\cite{Si1,Si2},
and discuss the prospect of experimentally measuring spin
transport as well as the status of some on-going spin
injection experiments in the cuprates\cite{Goldman,Venkatesan,Yeh,Lee}.

\section{What is needed to probe spin-charge separation}

Spin-charge separation is defined in terms of the excitations
of a many-electron system.
In essence, it says that A) there are two types of elementary 
excitations and B) the quantum numbers are such that, one
elementary excitation has spin ${1 \over 2}\hbar$ and charge $0$
(``spinon'') while the other has spin $0$ and charge $e$ (``holon'').
More precisely, imagine we have solved all the many-body eigenstates
of the 10$^{\rm 23}$ or so electrons.
Consider the many-body
excited states whose energies are not very far above the ground
state energy.
Spin-charge separation describes the situation when
it is necessary to introduce an elementary object 
with spin ${1 \over 2}\hbar$ and charge $0$,
{\it and} another with spin $0$ and charge $e$,
to reproduce the wavefunctions of these
low-lying many-body excited states  from the ground state
wavefunction $|{\rm gs}>$.
Namely, the many-body excited states have the general
form $ [ A_{\rm holon}^{\dagger} ]^l [ A_{\rm holon} ]^l 
[ A_{\rm spinon}^{\dagger} ]^m [ A_{\rm spinon} ]^n | {\rm gs} >$,
where $A_{\rm holon}^{\dagger}$ ($A_{\rm holon}$)
and $A_{\rm spinon}^{\dagger}$ ($A_{\rm spinon}$)
create (annihilate) a holon and a spinon, respectively.
In particular, the many-body states in the purely
spin and charge sectors, 
$|{\rm excited ~ states~I}>$ and 
$|{\rm excited ~ states~II}>$,
have to be  separately constructed from the ground state,
\begin{eqnarray}
|{\rm excited ~ states~I}> \sim  
[ A_{\rm spinon}^{\dagger} ]^n [ A_{\rm spinon} ]^m 
| {\rm gs} > \nonumber\\
|{\rm excited ~states~II}> \sim 
[ A_{\rm holon}^{\dagger} ]^n [ A_{\rm holon} ]^n 
| {\rm  gs} > 
\label{holons}
\end{eqnarray}
This definition parallels that for a Fermi liquid,
where only a single species of elementary excitations
is needed.
Introducing $A_{\rm qp}^{\dagger}$ ($ A_{\rm qp}$)
which creates (annihilates) a quasiparticle
with both spin ${1 \over 2}\hbar$ and charge $e$,
we can write for all the low-lying many-body excited states,
\begin{eqnarray}
|{\rm excited~states}> 
~\sim ~ [ A_{\rm qp}^{\dagger} ]^n 
[ A_{\rm qp} ]^n |{\rm gs} > \nonumber\\
{\rm for~a~Fermi~liquid}
\label{quasiparticle}
\end{eqnarray}
In a Fermi liquid, Landau parameters are also needed to specify
the residual interactions between these quasiparticles
(as well as to create collective excitations out of the 
quasiparticles). Similarly, in a spin-charge separated metal,
there are also parameters characterizing the
residual interactions among the spinons and holons.

Following the above definition, 
we can now specify the basic
elements of a signature of spin-charge separation.
It should not only show the existence of two kinds
of elementary excitations, but also provide the quantum
numbers of these excitations. Namely, we need to know
that one type of elementary excitations carry spin 
but no charge, while the other carry charge but no spin.

We comment in passing on the angle-resolved photoemission
spectroscopy (ARPES), which has been extensively discussed
in the literature in the context of spin-charge separation.
In ARPES, a physical electron - containing both spin
${1 \over 2}\hbar$ and charge $e$ of course - is ejected.
One doesn't know {\it a priori} whether any ARPES peak results
from a) a convolution involving a coherent spinon, a coherent
holon or both; b) a convolution involving other objects of
exotic quantum numbers; or c) simply a quasiparticle-like
excitation. From this perspective, ARPES does not
directly tell the quantum numbers of the elementary
excitations and, hence, doest not directly probe spin-charge
separation. Further discussions along this line can be
found in Ref. \cite{Rabello}.

\section{Probing spin-charge separation using spin transport}

Such a proposal was made a few years ago\cite{Si1,Si2}.
The basic idea is as follows.
Consider a spin current which will be generated by accelerating spins,
and charge current generated by accelerating charges.
We can infer that spin-charge separation exists if the
carriers for the two currents are two separated excitations.
Similarly, we can infer about the absence of spin-charge separation
if the carriers for the two currents actually correspond to the 
same excitation. Our proposal is that, we can determine which is
the case by comparing the temperature dependence of the spin resistivity
with that of the electrical resistivity.

To see this, we first note that the spin resistivity can be defined
in parallel to electrical resistivity. An electrical 
current, $J$, is established in the
steady state when charges are accelerated by an electric field, $E$.
The electrical resistivity is of course defined by the linear-response
ratio
\begin{eqnarray}
\rho = E / J 
\label{rho}
\end{eqnarray}
Similarly, a spin current,
$J_M$, will be established when
spins are accelerated by a magnetic field gradient,
$\nabla(-H)$. The ratio
\begin{eqnarray}
\rho_{spin} = \nabla(-H) / J_M
\label{rho_spin}
\end{eqnarray}
defines the spin resistivity. In a metal, the electrical resistivity 
$\rho$ is proportional to the transport relaxation rate $1/\tau_{tr}$,
which is the decay rate of the charge current. Similarly, $\rho_{spin}$ 
is proportional to the spin transport relation rate $1/\tau_{tr,spin}$,
the decay rate of the spin current.

In a spin-charge separated metal, spin current and charge current
are carried by different elementary
excitations. It then follows that the scattering processes
and the corresponding scattering phase space are in general
different for the decay
of spin current and decay of charge current. Therefore, the
two resistivities will have different temperature dependences.
We have calculated spin resistivities in models for spin-charge
separation\cite{Si1}, with results which indeed follow the 
above general conclusions. 
One model is the Luttinger liquid\cite{Voit} in which $\rho_{spin}
\propto T^{\alpha_{spin}}$ and $\rho \propto T^{\alpha_{charge}}$,
where the difference in the powers $\alpha_{spin} - \alpha_{charge}$ 
is non-zero and interaction dependent (for the one dimensional
Hubbard model $ 0 < \alpha_{spin} - \alpha_{charge} < 2$).
The other model is the U(1) gauge theory of the t$-$J
model\cite{Gauge}, in which $\rho_{spin} \propto T^{4/3}$ while $\rho
\propto T$.

Consider next the case without spin-charge separation. Here 
the same quasiparticle-like excitations carry both the spin
current and charge current. Any scattering process which
causes the decay of one current will necessarily also lead to
the decay of the other current. The two resistivities will then
have the same temperature dependences, under three conditions;
all these conditions are satisfied by at least the optimally
doped cuprates. The conditions are a) spin fluctuations are not
dominated by the ferromagnetic component; b) Fermi surface
is large; and c) inelastic scattering dominates over elastic
scattering. The conditions a) and b) have to do with the fact
that in establishing a spin current the spin up and spin down
excitations move in opposite directions. This leads to 
a matrix element for the decay of spin current that is slightly
different from its counterpart for the decay of charge current.
Conditions a) and b) guarantee that this difference in the matrix
elements leads only to a difference in
the numerical prefactors of the two resistivities,
but not in their temperature dependences.
Condition c) has to do with the possible fluctuating 
conductivities coming from fluctuating collective modes.
In the clean limit specified by condition c),
any fluctuating conductivity is better thought
of as a drag contribution; a very general gauge invariance
argument exists which guarantees that its
temperature dependence is the same as that of the 
corresponding Boltzmann contribution. Unlike for a) and b),
which are necessary, condition c) is sufficient but is in most
cases not necessary. From neutron scattering results, we know
that condition a) is satisfied for all cuprates.
Condition b) is satisfied at least for the optimally doped
cuprates, as can be inferred from the ARPES results.
Finally, from charge transport data 
we can infer that condition c) is satisfied 
for most cuprates.

In short, the spin resistivity can be used to test spin-charge 
separation. Over the temperature range 
where the electrical resistivity $\rho$ is linear in temperature,
a non-linear temperature dependence
of spin resistivity $\rho_{spin}$ would imply spin-charge separation
while a linear temperature dependence of $\rho_{spin}$ 
signals the absence of spin-charge separation.

\section{\bf Experimental measurement of spin resistivity}

While 
it is not easy to measure spin currents directly, 
the linear-response spin resistivity can be related to the spin
diffusion constant $D_s$ through the Einstein relation:
\begin{eqnarray}
\rho_{spin} = {1 / ( \chi_s D_s)}
\label{spin-diffusion}
\end{eqnarray}
where $\chi_s$ is the uniform spin susceptibility.
There are a number of experimental techniques
that can in principle be used to measure spin diffusion.
One feasible means for the cuprates seems to be
the spin-injection-detection technique\cite{Johnson}. 

The illustrative set-up can be found in 
Ref. \cite{Johnson}, and also in
Refs. \cite{Si2,Merrill}.
Basically, a cuprate is in contact with either one or two
ferromagnetic metal(s) (FM1 and in the latter case FM2 as well).
An electrical current (I)
is applied across the FM1-cuprate interface,
injecting spins into the cuprates.
In a steady state, the spatial distribution of the 
injected magnetization depends on the spin diffusion
of the cuprates.
The injected magnetization can be detected either by measuring
the voltage induced across the cuprate-FM2 interface,
or some alternative means.

To assess the feasibility of this kind of experiments, 
we need to estimate the spin diffusion length of
the cuprates. For the normal state, we can relate
the relaxation time for the total spin, $T_1$,
to the relaxation time for the spin current, $\tau_{tr,spin}$,
as follows,
\begin{eqnarray}
{ 1 / T_1} \approx \lambda_{so}^2 {( 1 / \tau_{tr,spin})}
\label{t1}
\end{eqnarray}
where $\lambda_{so}$ is the dimensionless spin-orbit coupling constant,
which for the cuprates is of the order of $0.1$.
Note that this relationship is generally valid independent
of the precise interactions responsible for the decay of the spin current
and spin; $\lambda_{so}$ measures the fraction
of the interaction processes which lose the total spin.
Combining Eq. (\ref{t1}) with the defining expression
for the spin diffusion length, $\delta_{spin} = \sqrt{2 D_s T_1}$,
we derive the following relationship between the spin diffusion length 
and spin transport mean free path,
\begin{eqnarray}
\delta_{spin} \approx { ( 1 /  \lambda_{so} )} l_{tr,spin}
\label{delta-spin}
\end{eqnarray}
From the above we have estimated the spin diffusion length
in the normal state to be in the range of a thousand $\AA$ to
a micron\cite{Si1}, long enough for
spin-injection-detection experiments. 

The recent spin injection experiments
in the cuprates\cite{Goldman,Venkatesan,Yeh,Lee}
became possible when perovskite manganites were used as
the ferromagnetic layer(s).
Presumably such manganite-cuprate heterostructures have 
much cleaner interfaces, since the two materials are 
structurally similar.
So far the experiments are restricted to the
superconducting cuprates.
They have raised many interesting theoretical
questions of their own, such as the physics
of Andreev reflection\cite{Zhu,Zutic,Kashiwaya,Merrill}
and the combined effects of Andreev reflection and 
single-particle transport on spin-injection
characteristics\cite{Merrill}
involving a ferromagnetic metal and 
a $d-$wave superconductor.

From the perspective of probing spin-charge separation,
it is necessary to extend these experiments to the
normal state and to carry out quantitative spin transport
measurements. Note that only the linear response regime
is needed for this purpose. 
As discussed in the previous section, what is needed is a plot
of spin resistivity as a function of temperature. There are
at least four possible ways to experimentally implement
such a plot. 
a) The most rigorous is to extract the spin diffusion
constant $D_s$ from experiments which, through Eq.
(\ref{spin-diffusion}), provides the spin resistivity directly;
b) Another possibility is to extract the spin diffusion length 
$\delta_{spin}$ from experiments.  Eq. (\ref{delta-spin}) implies
that $\delta_{spin}$ is proportional to $l_{tr,spin}$ which,
in  turn, is proportional to the spin conductivity. 
The temperature dependence of ${1 \over \delta_{spin}}$ 
is then the same as that of $\rho_{spin}$;
c) Along a similar line, it was shown\cite{Si2} that
the temperature dependence of $\rho_{spin}$ can be
directly inferred from the spin-dependent voltage $V_s$
in a spin-injection-detection experiment.
When the sample thickness $d$ is small compared to
$\delta_{spin}$, $\rho_{spin}$ is proportional to 
$I/ (\chi_s V_s)$ where $I$ is the injection current.
In the opposite limit of $d \gg \delta_{spin}$,
$\rho_{spin}$ has the same temperature dependence
as $\ln (I/\chi V_s)$;
d) Finally, from Eq. (\ref{t1}), $1/T_1$ is proportional
to the spin transport relaxation rate $1/\tau_{tr,spin}$
which in turn is proportional to the spin resistivity.
The temperature dependence
of $1/T_1$ is then the same as that of $\rho_{spin}$.
There are experimental techniques - such as ESR - which 
can measure $1/T_1$ (or $1/T_2$) directly. Following 
Eq. (\ref{t1}) we estimate the ESR linewidth to be in
the range of $3-30$GHz.


This work has been supported by NSF Grant No. DMR-9712626,
Robert A. Welch Foundation, Research Corporation,
and Sloan Foundation.

\end{document}